\documentclass{article}
\usepackage{spconf,amsmath,graphicx,array}
\title{IMPROVING AUDIO CAPTIONING USING SEMANTIC SIMILARITY METRICS}
\name{Rehana Mahfuz, Yinyi Guo, Erik Visser}%\thanks{Thanks to XYZ agency for funding.}}
\address{Qualcomm Technologies, Inc.}
\address{\{rmahfuz, yinyig, evisser\}@qti.qualcomm.com}
\begin{document}
%\ninept
%
\maketitle
\begin{abstract} %100-150 words
Audio captioning quality metrics which are typically borrowed from the machine translation and image captioning areas measure the degree of overlap between predicted tokens and gold reference tokens. In this work, we consider a metric measuring semantic similarities between predicted and reference captions instead of measuring exact word overlap. We first evaluate its ability to capture similarities among captions corresponding to the same audio file and compare it to other established metrics. We then propose a fine-tuning method to directly optimize the metric by backpropagating through a sentence embedding extractor and audio captioning network. Such fine-tuning results in an improvement in predicted captions as measured by both traditional metrics and the proposed semantic similarity captioning metric.
%In the field of audio captioning, metrics established to measure the quality of the caption are metrics borrowed from either machine translation or image captioning. Since machine translation metrics look for intersection between exact words or tokens, we consider a metric which captures semantic similarities by computing the cosine distances between sentence embeddings of the generated and reference captions. To compare metrics, we evaluate their abilities to capture similarities among captions corresponding to the same audio file. Finally, we propose a fine-tuning method to directly optimize the metric by backpropagating through the sentence embedding network in tandem with the audio captioning network. Such fine-tuning results in an improvement in captioning metrics as well as in the metric that captures semantic similarity.
\end{abstract}
\section{Introduction}
\label{sec:intro}

The ability to understand the world around us using sound can afford edge AI devices a new level of context awareness. While a lot can be perceived by understanding images and videos, the auditory scene includes complementary information that cannot be captured by a camera without a microphone. For example, a CCTV camera may capture a video of burglars, but only the microphone can capture elements such as the voices of intruders, no matter where the camera is pointed. %Additionally, a microphone may have a wider range of coverage than a camera. 
While a lot of information can be retrieved from images and videos, an auditory scene captured by a microphone includes complementary information that cannot be captured in a visual field.
For example, the camera will only capture the burglar breaking glass or picking the lock or breaking the door if the camera is pointed in the direction of the door or window. However, a microphone can simultaneously capture sounds of glass breaking, pounding on the door, squeaking of lock picking, jingling of keys, footsteps, voices and much more. Hence, understanding of audio is crucial to developing automated awareness of context.

The first step towards developing audio understanding was audio tagging, which involves detecting the occurrence of any common sounds from a finite set of sounds. Given that this has been attempted with some success, more can be achieved by being able to describe the sounds using language natural to humans. 
%Because generation of flowing text is possible, effort has been made to generate text to describe sounds. Generating audio captions in natural language also affords us the ability to describe sounds which cannot be labelled since they don't fit into any of the common categories. 
Such descriptions are natural for human-machine interface applications and can be readily analyzed with modern Natural Language Processing tools \cite{Ravi_paper}.
Generating audio captions in natural language also enables description of sounds which cannot be directly tagged with pre-defined labels since they don't fit into any of the common categories.
For example, falling of rocks is not a commonly recorded sound, and hence cannot be described well by audio tagging. The best that audio tagging can do is label it as a ``thunk". More justice can be done to the description of such a sound if an audio captioning engine is able to describe the sound as ``falling of hard objects", or ``collision of solid objects".

%The effort to caption audio began with the availability of datasets of audio with human-generated captions.
A number of audio datasets with human-generated captions are available in the literature \cite{audiocaps, clotho} and models have been developed to predict natural language audio captions \cite{rl_finetune, dcase_mei}.
%, dcase_mei, dcase_primus, dcase_kouzelis}.
As in machine translation, one method to evaluate the quality of the generated caption has been to calculate overlap of n-grams between the predicted caption and the reference captions, in the form of precision, F-score or recall, termed as BLEU \cite{bleu}, ROUGE \cite{rouge} and METEOR \cite{meteor} respectively. Demanding exact overlap between n-grams is a harsh criterion, and often overlooks synonyms or words/phrases with similar meanings. While the metric METEOR tries to alleviate this problem by accounting for synonyms and words with common stems, such leniency may not be enough to account for words that convey similar meanings. Other audio captioning metrics are adapted from image captioning. Instead of considering n-grams of exact words or their stems/synonyms, the CIDER \cite{cider} metric looks at n-grams of Term Frequency-Inverse Document Frequency (TF-IDF)s \cite{tfidf}. Specifically, CIDER calculates cosine similarities between n-grams of TF-IDFs, and averages these for $n=1-4$. This is a significant step in moving away from demanding exact matches. Another image captioning metric SPICE \cite{spice} creates a scene graph from the set of reference captions, and tries to find an overlap between the scene graphs of the candidate caption and that of the reference captions. The limitation of this method is that it looks for exact matches between components of the scene graph. 
%Hence, we look into metrics which capture semantic similarities by leveraging word/sentence embeddings. Another way to judge semantic similarity is to consider the output of a model trained with the Natural Language Inference \cite{nli} task, which determines if two sentences are in entailment of, in contradiction to, or are neutral with respect to each other. %However, the bar is quite high for two sentences to be labelled as being in entailment, and different captions which mean almost the same thing don't always reach this bar.
In this paper, we instead propose metrics which capture semantic similarities between natural language descriptions leveraging word or sentence embeddings for the purpose of audio captioning model training and evaluation. This paper is organized as follows. Section 2 covers some related work. Section 3 describes the procedure, Section 4 explains results, and Section 5 provides a conclusion.

\section{Related Work}
\label{sec:related_work}
A typical audio captioning model \cite{rl_finetune,  dcase_mei} %\cite{xu_survey, xinhao_survey} 
includes an encoder, which extracts meaningful audio features containing discriminative information relevant to the audio scene, followed by a decoder which takes those features as input, and generates text. For the encoder, a network trained with an audio understanding task such as Sound Event Detection \cite{sed}, Acoustic Scene Classification \cite{asc} or Audio Tagging \cite{at} is used. This may be an RNN, a CNN or a CRNN. The decoder is a auto-regressive component, such as an RNN or a Transformer decoder \cite{transformer} which generates token probabilities.
The cross-entropy loss between one-hot encoded vectors of token IDs is generally used, as shown in Equation \ref{eq:CEloss}. Some work has also been attempted in directly optimizing the captioning metric using reinforcement learning, by using the score as a reward \cite{rl_finetune}.
%Describe audio captioning methods. Explain Figure 1(a)
%Describe all metrics
%Metrics established to evaluate the quality of audio captions can be classified into two types. Metrics borrowed from machine translation include BLEU, ROUGE and METEOR. Metrics adapted from image captioning include SPICE, CIDER and their average, SPICE. BLEU\_n \ref{bleu} measures the precision of the candidate sentence with respect to the reference sentence by matching n-grams. ROUGE-n\cite{rouge-n} is a similarly calculated recall measure. ROUGE-L is the recall when looking for the longest common subsequence match. The METEOR\cite{meteor} metric broadens its matching criterion from demanding exact matches to accommodating common word stems and synonyms. It calculates the harmonic mean of the precision and recall, where the precision is weighed nine times as much as the recall. The SPICE metric does xyz

Apart from the metrics described in Section \ref{sec:intro}, another metric FENSE \cite{fense} uses sentence embeddings to capture semantic similarities between captions. Additionally, it penalizes this score by using the output of a neural network trained to detect frequently occurring errors in captioning systems.
In our work, we leverage a sentence embedding based similarity metric for the purpose of fine-tuning an audio captioning model, which to our knowledge has not been reported before.
%We leverage this idea in our work to fine-tune our audio captioning model.

\section{Procedure}
\label{sec:procedure}

\subsection{Baseline system}
\label{ssec:baseline_system}

\begin{figure*}[htb]

%\begin{minipage}[b]{1.0\linewidth}
%  \centering
%  \centerline{\includegraphics[width=8.5cm]{generic_baseline.PNG}}
%  \vspace{2.0cm}
%  \centerline{(a) Established baseline system}\medskip
     %\centering
%     \begin{subfigure}%[b]{0.3\textwidth}
%      %   \centering
%         \includegraphics[width=16cm]{generic_baseline.PNG}
%         \caption{Established baseline system}
%         \label{fig:est_baseline}
%     \end{subfigure}
%     \begin{subfigure}%[b]{0.3\textwidth}
%      %   \centering
%         \includegraphics[width=16cm]{our_baseline.PNG}
%         \caption{Our baseline system}
%         \label{fig:our_baseline}
%     \end{subfigure}
%     \begin{subfigure}%[b]{0.3\textwidth}
      %   \centering
         \includegraphics[width=16cm]{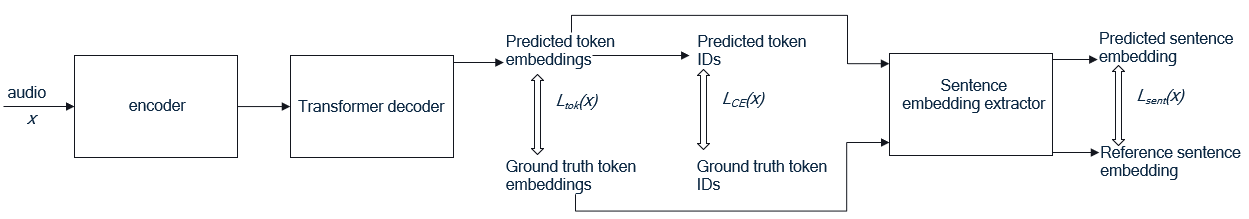}
         \caption{Our fine-tuning approach.}
         \label{fig:finetune}
%     \end{subfigure}

%\includegraphics[width=16cm]{generic_baseline.PNG}
%  \caption{Established baseline system}
%  \label{fig:est_baseline}
%\includegraphics[width=16cm]{our_baseline.PNG}
%  \caption{(Our baseline system}
%  \label{fig:our_baseline}
%\includegraphics[width=16cm]{finetuning.PNG}
%  \caption{(Our fine-tuning approach}
%  \label{fig:finetune}
  
%\end{minipage}
%
%\begin{minipage}[b]{1.0\linewidth}
%  \centering
%  \centerline{\includegraphics[width=8.5cm]{our_baseline.PNG}}
%  \vspace{1.5cm}
%  \centerline{(b) Our baseline system}\medskip
%  \label{fig:our_baseline}
%\end{minipage}
%\hfill
%\begin{minipage}[b]{1.0\linewidth}
%  \centering
%  \centerline{\includegraphics[width=8.5cm]{finetuning.PNG}}
%  \vspace{1.5cm}
%  \centerline{(c) Our fine-tuning approach}\medskip
%  \label{fig:finetune}
%\end{minipage}
%
%\caption{Incremental development of our approach.}
%\label{fig:1}
%
\end{figure*}

The baseline system has an encoder-decoder structure. The encoder is the CNN10 PANN \cite{pann} pre-trained for audio tagging. For the decoder, we use a stack of two transformer decoder layers with four heads. % and $gelu$ \cite{gelu} activation.
We train this decoder to generate token embeddings of the caption. These token embeddings are further projected into a space whose dimension is equal to the size of the vocabulary, so that the prediction can be expressed as a one-hot encoded vector. As shown in Equation \ref{eq:CEloss}, the training objective is to minimize the cross-entropy loss between the one-hot encoded vector representations of the reference caption $p(x)$ and of the predicted caption $q(x)$, where $x$ is one of the many audio clips from the collection of audio clips $X$.
%In an effort to make the token embeddings more accurate, we also tried adding to the loss the cosine distance between token embeddings of the predicted caption $R(x) = [r_1(x), r_2(x), ..., r_{n_{ref}}(x)]$ and of the reference caption $S(x) = [s_1(x), s_2(x), ..., s_{n_{ref}}(x)]$, where $n_{ref}(x)$ denotes the number of tokens in the reference caption of audio clip $x$. This additional loss term is shown in Equation \ref{eq:token-loss}.
This network has about 8M parameters if 128-dimensional word2vec \cite{word2vec} embeddings are used, and about 14M trainable parameters if 384-dimensional BERT \cite{bert} embeddings are used. To evaluate results, in addition to metrics introduced in Section \ref{sec:related_work}, we also use the Word Mover's Distance (WMD) \cite{wmd}, which establishes a correspondence between similar words in two sentences, and calculates the distance that words of one sentence would have to travel to reach the other sentence.
\begin{equation}
\label{eq:CEloss}
L_{CE}(X) = -\sum_{x \in X} p(x)log(q(x))
\end{equation}

%\begin{align}
%\label{eq:token-loss}
%L_{tok}(X) = \sum_{x \in X} 
%\frac{1}{n_{ref}(x)}\sum_{i=1}^{n_{ref}(x)} 1-%\frac{r_i(x).s_i(x)}{||r_i(x)|| ||s_i(x)||}
%\end{align}

\subsection{Fine-tuning to optimize metric}
\label{ssec:finetune}
%Given that the sentence embedding of the caption captures the gist of its meaning in one vector, the cosine similarity between the candidate and reference captions is a meaningful measure of the quality of the caption. 
The use of sentence embeddings to capture the meaning of an entire caption in one vector has been proposed \cite{fense}. The cosine similarity between candidate and reference caption sentence embeddings is thus a meaningful measure of candidate caption quality. 
In an effort to optimize this metric directly using backpropagation, we append the sentence embedding network to the audio caption generation network, as shown in Figure \ref{fig:finetune}. Specifically, the token embeddings generated by the decoder just before the token IDs are directly input into the sentence embedding extractor. For fine-tuning the trained network, we add to the cross-entropy loss the cosine distance between the sentence embedding of the generated caption $t(x)$ and the sentence embedding of the reference caption $u(x)$. This loss term is shown in Equation \ref{eq:sent-loss}.
%We also experiment with only including $L_{CE}(x)$ and $L_{sent}(x)$ in the loss function.
Hence, the weights of the audio captioning network are directly optimized to minimize the distance between sentence embeddings, and to consequently make the generated caption closer in meaning to the reference caption. Weights of the sentence embedding network are frozen during fine-tuning.
%No other weights are frozen, including the weights of the PANN encoder.
Performing such fine-tuning does not affect the size of the model used during inference, since inference stops once predicted token IDs are obtained, as shown in Figure \ref{fig:finetune}.

\begin{align}
\label{eq:sent-loss}
L_{sent}(X) = \sum_{x \in X} 1-
\frac{t(x).u(x)}{||t(x)|| ||u(x)||}
\end{align}

We chose the Sentence-BERT \cite{sbert} sentence embedding network, which was trained with the objective of classifying if two sentences are in entailment of, in contradiction to, or neutral with respect to each other. It accepts token-level BERT \cite{bert} embeddings of the sentence, passes it through a version of BERT which was fine-tuned for sentence embedding generation, and then performs mean pooling to generate a 384-dimensional representation of the sentence. The twelve-layer version of BERT was reduced to six layers by keeping only every second layer.
Along these lines, we also define the audio captioning metric SBERT\_{sc} to be the cosine similarity between Sentence-BERT embeddings of the candidate and reference captions.
For the audio captioning model, the choice of token-level embeddings has to be BERT, so that the token-level embeddings generated before the one-hot encoded vectors representing token IDs can be directly input into Sentence-BERT to enable end-to-end backpropagation.

\subsubsection{Evaluating metrics}
\label{sssec:comparing_metrics}
To evaluate the effectiveness of audio captioning metrics, we devised a simple method. Given a database of sounds along with their human-generated reference captions, we evaluate the scores between all pairs of captions. A good metric will yield a high score for captions describing the same audio file, and a low score for captions describing audio files with different content. For a candidate metric, we average its score over all pairs of captions that describe the same audio file, henceforth referred to as \textit{average-over-similar}, abbreviated as $AS$, and over all pairs of captions that describe distinct audio files, henceforth referred to as \textit{average-over-distinct}, abbreviated as $ADs$. However, captions for audio files with similar content may be similar. %provide example of clotho_file_20091225.rain.01.wav_0.npy and clotho_file_080809_05_FontanaKoblerov.wav_0.npy
Hence, from $ADs$, we eliminate scores of all pairs of captions which correspond to distinct files with overlapping audio tags. We term this score as \textit{average-over-different}, abbreviated as  $ADf$.

For each metric, we normalized the scores to be between 0 and 1.
For the SPICE metric, since all reference captions are needed to calculate the scene graph, pairwise scores do not make sense. Hence, the score of each candidate caption was computed with respect to groups of captions, where each group corresponds to one audio file.

% We also propose the CIDER-WE metric, which is CIDER, but using word embeddings.
% Try: CIDER using BERT embeddings

\subsection{Implementation Details}
\label{ssec:implementation_details}

\subsubsection{Datasets}
\label{ssec:datasets}

For evaluating metrics, we used the validation set of the AudioCaps \cite{audiocaps} dataset. This dataset consists of 49,838 audio files for training with one human-generated caption per file. It also contains 495 validation audio files and 975 test audio files, all of which are annotated with five human-generated captions per audio file. The audio files are of 10 s in length, and are sourced from the AudioSet database.

For training the audio captioning models, we used the AudioCaps dataset for initial training, followed by further training with the Clotho \cite{clotho} dataset. The Clotho dataset contains audio files from the Freesound database with durations ranging from 15 s to 30 s, where each audio file is annotated with five different human-generated captions. The train, validation and test splits have 3839, 1045 and 1045 audio files respectively.

\subsubsection{Training Details}
\label{ssec:trn_details}

For both the baseline approach and the fine-tuning approach, the networks were trained for 30 epochs with a batch size of 32. The best epoch was picked, based on the validation loss. %The Adam \cite{adam} optimizer was used.
The learning rate started with 0.001, and decayed by a factor of 0.1 every 10 epochs. For decoding, we only used greedy decoding.

To train the model for the Clotho dataset, we used the best AudioCaps model after fine-tuning using Sentence-BERT.
\section{Results}
\label{sec:results}

\subsection{Metric Evaluation}
\label{ssec:metric_eval_results}

%>{\centering}p{2.5cm}
\begin{table}[ht]
\centering
%\begin{tabular}{||c c c c c||} 
\begin{tabular}{||p{0.15\linewidth} p{0.13\linewidth} p{0.13\linewidth} p{0.13\linewidth} p{0.18\linewidth}||}
 \hline
 %Metric & $AoS$ & $AoDs$ & $AoDf$ & \multicolumn{2}{l}{$AoS-AoDf$} \\ 
 Metric & $AS$ & $ADs$ & $ADf$ & $AS-ADf$\\ 
 %&$\_same$ & $\_distinct$ & $\_different$ & $\_same-avg\_over$ \\
 %&&&&$\_different$\\ [0.5ex]
 \hline\hline
 CIDER & 0.0788 & 0.0014 & 0.0006 & 0.0783\\ 
 SPICE & 0.4553 & 0.0295 & 0.0275 & 0.4278\\
 SBERT\_sc & \textbf{0.6410} & \textbf{0.2172} & \textbf{0.1918} & \textbf{0.4492}\\  [1ex]
 \hline
\end{tabular}
\caption{Comparison of different metrics.}
\label{table:metric-comp-result}
\end{table}

From Table \ref{table:metric-comp-result}, we observe that the value of $AS-ADf$ is the highest for SBERT\_sc, declaring it as the metric most capable of differentiation between captions belonging to unrelated audio, and of establishing similarities between descriptive captions of the same audio. Hence we decided to optimize this metric via backpropagation.
The CIDER score seems low because we normalized it to be between 0 and 1, for fair comparison with the other metrics, which are also scaled between 0 and 1.

%From Table \ref{table:metric-comp-result}, we observe that the value of $avg\_over\_same - avg\_over\_different$ is the highest for CIDER, declaring it as the metric most capable of differentiation among captions belonging to unrelated audio, and of establishing similarities between descriptive captions of the same audio. However, since it is hard to optimize for CIDER via backpropagation, we chose to optimize the next-best metric, SBERT\_sc.

\subsection{Result of Finetuning}
\label{ssec:result_of_finetuning}

\begin{table*}[h]
\centering
%\begin{tabular}{||c c c||} 
\begin{tabular}{||p{0.15\linewidth} p{0.06\linewidth} p{0.45\linewidth} p{0.09\linewidth} p{0.09\linewidth}||}
 \hline
  && Captions & SBERT\_sc & SBERT\_sc\\ [0.5ex] 
%  \hline
  & & & before FT & after FT \\
%  \hline
 \hline\hline
 Predicted caption & w/o FT & a woman speaks and a baby cries &&\\ 
 & w/ FT & a woman speaks and laughs &&\\ 
 Reference captions  & 1 & a female speech and laughing with running water &0.360 & \textbf{0.661}\\
 & 2 & ocean waves crashing as a woman laughs and yells as a group of boys and girls scream and laugh in the distance & 0.353 & \textbf{0.462}\\
 & 3 & a woman yelling and laughing as a group of people shout while ocean waves crash and wind blows into a microphone & 0.417 & \textbf{0.637}\\
 & 4 & woman screaming and yelling to another person with loud wind and waves in background & 0.472 & \textbf{0.505}\\
& 5 & a woman yelling and laughing as ocean waves crash and a group of people shout while wind blows into a microphone & 0.420 & \textbf{0.631}\\
 \hline

 Predicted caption & w/o FT & wind blows strongly and a man speaks &&\\
 & w/ FT & wind blows strongly and a gun fires &&\\
 Reference captions & 1 & several very loud explosions occur & 0.283 & \textbf{0.445}\\
 & 2 & loud bursts of explosions with high wind & 0.450 & \textbf{0.590}\\
 & 3 & a loud explosion as gusts of wind blow & 0.538 & \textbf{0.629}\\
 & 4 & loud banging followed by one louder bang with some staticloud banging followed by one louder bang with some static & 0.273 & \textbf{0.353}\\
 & 5 & very loud explosions with pops and bursts of more explosions & 0.248 & \textbf{0.413}\\
 [1 ex]
 \hline
\end{tabular}
\caption{Examples of the outcome of fine-tuning (FT).}
\label{table:result_example}
\end{table*}

\begin{table*}[!ht]
\centering
%\begin{tabular}{||c c c c c c c c c c c c||} 

\begin{tabular}{||p{0.06\linewidth} p{0.055\linewidth} p{0.055\linewidth} p{0.055\linewidth} p{0.055\linewidth} p{0.055\linewidth} p{0.062\linewidth} p{0.044\linewidth} p{0.044\linewidth} p{0.072\linewidth} p{0.054\linewidth} p{0.06\linewidth}||} 
 \hline
 & BLEU\_1 & BLEU\_2 & BLEU\_3 & BLEU\_4 & ROUGE & METEOR & CIDER & SPICE & SBERT\_sc & WMD & FENSE-error\\ [0.5ex] 
 \hline\hline
 &&&&&AudioCaps&&&&&&
 \\ \hline
 w/o FT & \textbf{0.6612}&\textbf{0.4761}&\textbf{0.3207}&0.2020&\textbf{0.4558}&0.2159&0.5730 & 0.1575 & 0.5447 & 6.5078 & 0.4901\\ 
w/ FT &0.6486&0.4692&0.3204&\textbf{0.2080}&0.4554&\textbf{0.2178}&\textbf{0.5735} & \textbf{0.1664} & \textbf{0.5552} & \textbf{6.4503} & \textbf{0.4815}\\
  \hline
&&&&&Clotho&&&&&&
\\ \hline
w/o FT & \textbf{0.5253}&0.3196&\textbf{0.1974}&\textbf{0.1200}&\textbf{0.3573}&0.1569&0.3061 & 0.1049 & 0.4189 & 6.9616 & 0.6328\\
w/ FT & 0.5228&\textbf{0.3198}&0.1952&0.1183&0.3559&\textbf{0.1582}&\textbf{0.3097} & \textbf{0.1062} & \textbf{0.4262} & \textbf{6.8194} & \textbf{0.6182}\\ [1ex]
 \hline
\end{tabular}
\caption{Scores before and after fine-tuning (FT). Higher is better for all metrics except WMD and FENSE-error.}
\label{table:finetune-result}
\end{table*}

%Using the loss function $L_{CE}(x) + L_{tok}(x)$ did not result in a noticeable improvement compared to using the loss function $L_{CE}(x)$, which is why we decided to not include $L_{tok}(x)$ in the loss function while fine-tuning. Hence, 
The baseline before fine-tuning used the $L_{CE}(x)$ loss, and fine-tuning was performed with the $L_{CE}(x) + L_{sent}(x)$ loss.
From Table \ref{table:finetune-result}, we observe that the fine-tuning step using Sentence-BERT indeed results in an improvement using both image captioning metrics as well as sentence embedding based metrics. As illustrated in Table \ref{table:result_example}, after fine-tuning, the model generates captions whose SBERT\_sc are higher with respect to those of the reference captions, since the captions are closer in meaning to the reference captions after fine-tuning.  However, using translation metrics such as BLEU and ROUGE, the improvement is not obvious because the improvement is in semantic similarity and not in exact word overlap. The error, as detected by the error detector of the FENSE metric, also decreases. The large error is caused by incorrect error detection in the \textit{missing verb} category, in captions such as ``a spray is released", ``a man is speaking", ``a dog is growling", ``water is rushing by" and ``a door is closed".
%The only translation metric that shows a clear improvement is METEOR, which is not surprising, since that is the only one which accommodates matching of word stems and of synonyms. This suggests that fine-tuning the model does not necessarily cause the model to generate words or n-grams which are exact matches of the reference.
The Word Mover's Distance (WMD) also decreases after fine-tuning, since the distance needed to travel from word embeddings of one caption to reach word embeddings of the other caption decreases.

%\section{Discussion}
%\label{sec:discussion}
%We find that metrics which use word/sentence embeddings are more effective in identifying the semantic closeness among audio captions.
%Another thing to consider while choosing to use a sentence-embedding-based fine-tuning method was that the tokenizer of the token embeddings that the model generates has to be the same as the tokenizer used by the sentence embedding network. Hence, it was an obvious choice to use for the audio captioning network the same word embeddings used by the sentence embedding method, which was 384-dimensional BERT embeddings. This caused the number of trainable parameters to increase to 14M from 8M if 128-dimensional word2vec embeddings were used. In the interest of controlling the network size, another trade-off is the choice of the encoder. Among PANNs, using CNN14 or another larger PANN rather than CNN10 may have afforded us better performance at the cost of a larger number of trainable parameters.

%Of course, our result is subject to the effectiveness of the metric that is being used, which calls for more research to establish a credible metric.
\section{Conclusion}
\label{sec:conclusion}
We demonstrate that fine-tuning a vanilla audio captioning model with the objective of minimizing the cosine distance between sentence embeddings of the candidate and the reference captions results in higher caption quality. Future work will focus on exploring metrics that capture semantic similarities at a lower computational cost.
%In this work, we experimented with only the Sentence-BERT embedding generator. It may be valuable to compare the usefulness of alternate sentence embedding generators such as InferSENT \cite{infersent} or Universal Sentence Encoder \cite{use}, which have lesser parameters, and may be more tractable for using as a metric as well as for fine-tuning.

% To start a new column (but not a new page) and help balance the last-page
% column length use \vfill\pagebreak.
% -------------------------------------------------------------------------
%\vfill
%\pagebreak

%\vfill\pagebreak

%\section{REFERENCES}
%\label{sec:refs}

% References should be produced using the bibtex program from suitable
% BiBTeX files (here: strings, refs, manuals). The IEEEbib.bst bibliography
% style file from IEEE produces unsorted bibliography list.
% -------------------------------------------------------------------------
\bibliographystyle{IEEEbib}
\bibliography{refs}

\end{document}